# Generation and Dynamics of Quantized Vortices in a Unitary Fermi Superfluid

Aurel Bulgac[1], Yuan-Lung (Alan) Luo[1], Piotr Magierski[2,1], Kenneth J. Roche[3,1], and Yongle Yu[4]

[1]*Department of Physics, University of Washington, Seattle, WA 98195–1560, USA*
[2]*Faculty of Physics, Warsaw University of Technology, ulica Koszykowa 75, 00-662 Warsaw, Poland*
[3]*Pacific Northwest National Laboratory, Richland, WA 99352, USA and*
[4]*Wuhan Institute of Physics and Mathematics, CAS, 430071, Wuhan, P.R. China*

Superfluidity and superconductivity are remarkable manifestations of quantum coherence at a macroscopic scale. The dynamics of superfluids has dominated the study of these systems for decades now, but a comprehensive theoretical framework is still lacking. We introduce a local extension of the time-dependent density functional theory to describe the dynamics of fermionic superfluids. Within this approach one can correctly represent vortex quantization, generation, and dynamics, the transition from a superfluid to a normal phase and a number of other large amplitude collective modes which are beyond the scope of two-fluid hydrodynamics, Ginzburg-Landau and/or Gross-Pitaevskii approaches. We illustrate the power of this approach by studying the generation of quantized vortices, vortex rings, vortex reconnection, and transition from a superfluid to a normal state in real time for a unitary Fermi gas. We predict the emergence of a new qualitative phenomenon in superfluid dynamics of gases, the existence of stable superfluidity when the systems are stirred with velocities significantly exceeding the nominal Landau critical velocity in these systems.

Superfluidity has been put in evidence in a large range of systems: in various condensed matter systems, in pure and mixtures of liquid $^3$He and $^4$He, in nuclei and in neutron stars, in both fermionic and bosonic cold atoms in traps, and it is predicted to show up in dense quark matter as color superconductivity. The critical temperatures at which a system undergoes a phase transition from a superfluid to a normal phase have a range approaching 20 orders of magnitude. Surprisingly the superfluid properties of many of these systems are qualitatively similar and can be described within a common framework. In spite of turning a century old in 2011 and the efforts of many generations of physicists, superfluidity still holds many of its secrets locked. Landau [1], following ideas of Tisza [2, 3], developed perhaps one of the most successful phenomenological models of these systems, the two-fluid hydrodynamics, by assuming that superfluid liquid $^4$He behaves below the critical temperature like a mixture of normal and superfluid components. A little more than a half a century ago Onsanger [4] and Feynman [5] predicted a remarkable property of superfluids, their ability to sustain quantized vortices. Using the Ginzburg-Landau (GL) phenomenological theory [6], which is valid for temperatures close to the critical temperature, Abrikosov was able to predict that many vortices would form the celebrated triangular lattice [7]. With the advent of the Bardeen-Cooper-Schrieffer (BCS) theory [8] most of these phenomenological models and predictions for the fermionic superfluids acquired a firm microscopic underpinning. For a weakly interacting Bose gas in the superfluid phase one can derive an equation formally similar to the GL-equation, the Gross-Pitaevskii (GP) equation [9, 10]. With its help one can demonstrate the existence of vortices and of a vortex lattice in Bose superfluids at temperatures well below the critical temperature. Feynman has also conjectured that superfluids can exhibit what later became known as quantum turbulence [5, 11], a phenomenon believed to be linked to the reconnection of quantized vortex lines in a vortex tangle.

In spite of their spectacular success traditional theoretical approaches are inadequate in many respects, however, and they fail to correctly describe a large class of phenomena in superfluids.

On one hand, Planck's constant does not appear in the Landau's two-fluid hydrodynamics. At zero temperature its equations reduce to the following two equations:

$$\dot{n}(\mathbf{r},t) + \boldsymbol{\nabla}\cdot[\mathbf{v}(\mathbf{r},t)n(\mathbf{r},t)] = 0,$$
$$m\dot{\mathbf{v}}(\mathbf{r},t) + \boldsymbol{\nabla}\left\{\frac{m\mathbf{v}^2(\mathbf{r},t)}{2} + \mu[n(\mathbf{r},t)] + V_{ext}(\mathbf{r},t)\right\} = 0.$$

Here $n(\mathbf{r},t)$ is the number density, $\mathbf{v}(\mathbf{r},t)$ is the superfluid velocity, $\mu[n(\mathbf{r},t)]$ is the local chemical potential, and $V_{ext}(\mathbf{r},t)$ is an arbitrary external potential which might be applied to the system. Even though a quantized vortex is allowed to exist as a solution of these equations of motion, there is no intrinsic mechanism preventing quantized vortices from dynamically evolving into non-quantized vortices, as there is no intrinsic mechanism leading to the formation of a quantized vortex. Another drawback is the absence of Landau's critical velocity in this framework.

Because of that, there is no physically clear mechanism which would allow the transition from the superfluid to the normal phase under the action of some external agent, such as a fast moving piston for example.

On the other hand, the GL-equation is valid only close to and just below the critical temperature, when the superfluid parameter is small in amplitude. The mathematically similar in structure GP-equation, on the other hand, describes a Bose condensed system at temperatures much lower than the critical temperature. In both GL and GP approaches the mathematical equation to be solved has the form of a nonlinear Schrödinger equation

$$i\hbar\frac{\partial \Psi(\mathbf{r},t)}{\partial t} = -\frac{\hbar^2 \Delta \Psi(\mathbf{r},t)}{2M} + \mathcal{U}(|\Psi(\mathbf{r},t)|^2)\Psi(\mathbf{r},t) + V_{ext}(\mathbf{r},t)\Psi(\mathbf{r},t),$$

where $M$ is the mass of the boson or of the Cooper pair respectively. This equation describes only the spatio-temporal evolution of the superfluid order parameter and within this approach the quantized vortices and the Abrikosov triangular vortex lattice emerge naturally. However, if a superfluid is stirred with a velocity larger than the Landau's critical velocity [12], either a part of the system or the whole system can become normal and the order parameter vanishes in the corresponding region of space. This kind of transition from a superfluid to a normal phase cannot be captured within a GL/GP-approach.

The excitation of the so called Higgs mode of the pairing field and other related modes [13–15] are examples of when phenomenological approaches fail spectacularly, while the existence of such modes can be proven in exactly solvable models [16]. These modes are extremely slow, with frequencies much lower than the pairing gap, and they typically have a very large amplitude. It has been assumed always that slow modes never lead to any significant distortions of the quasiparticle distributions, namely that a local either Bose-Einstein or Fermi-Dirac equilibrium exists. It is this assumption of the existence of a local equilibrium that allowed previous authors to justify these phenomenological models in the limit of slow motion, and this assumption is the reason these approaches fail in the case of these modes.

We will focus here on the dynamics of the Unitary Fermi Gas (UFG) in order to address these crucial aspects and also to demonstrate the existence of new qualitative phenomena in a range of superfluid systems. The UFG became an object of intense theoretical and experimental study barely a decade ago (see reviews [17–21]) and over the years it proved to be a system with truly remarkable properties. When Bertsch formulated the Many-Body X Challenge in 1999 [22] neither compelling theoretical arguments nor experimental evidence existed which would point to either the stability or instability of a UFG. The first *ab initio* calculation of the UFG ground state properties performed in 2003 [23] and the first experimental realization of a quantum degenerate cloud of cold fermionic $^6$Li atoms in an atomic trap [24] were the first incontrovertible theoretical and experimental arguments in this direction. The subsequent observation of the vortex lattice provided the conclusive experimental proof that this system is indeed superfluid [25]. It became clear relatively recently that the UFG has the highest Landau critical velocity of any known superfluid [26, 27], as well as having a very high critical temperature.

We have used the Hohenberg-Kohn approach to Density Functional Theory (DFT) [28] and its time-dependent extension [29] in order to develop a highly accurate theoretical framework capable of explicitly treating the pairing correlations and the time-dependent (TD) phenomena in a UFG [30, 31]. This framework is known as the Time-Dependent Superfluid Local Density Approximation (TDSLDA). The case of a UFG is special, as dimensional arguments alone place strong limits on the form of the functional. Galilean invariance can be invoked to determine its dependence on currents, which emerge when time-dependent phenomena are considered. The solution of the TDSLDA equations in 3D and time, which look formally like TD selfconsistent Bogoliubov-de Gennes (BdG) equations, require a rather sophisticated use of leadership class supercomputers [32]. The numerical complexity of these equations was a serious stumbling block until now and only a single limited 2D solution of TDBdG equations has been attempted [33]. The full 3D solution of the TDSLDA equations amounts to numerically solving tens to hundreds of thousands of coupled TD non-linear Schrödinger equations. We will demonstrate here that within the time-dependent extension of DFT, when the quasiparticle wave functions $u_{n,\sigma}(\mathbf{r},t)$, $v_{n,\sigma}(\mathbf{r},t)$ satisfy the TDSLDA equations, all the limitations of the traditional approaches discussed above are eliminated. In the absence of a spin-orbit coupling these equations acquire the form

$$i\hbar\frac{\partial}{\partial t}\begin{pmatrix} u_\uparrow(\mathbf{r},t) \\ v_\downarrow(\mathbf{r},t) \end{pmatrix} = \begin{pmatrix} h_\uparrow(\mathbf{r},t) & \Delta(\mathbf{r},t) \\ \Delta^*(\mathbf{r},t) & -h^*_\downarrow(\mathbf{r},t) \end{pmatrix}\begin{pmatrix} u_\uparrow(\mathbf{r},t) \\ v_\downarrow(\mathbf{r},t) \end{pmatrix},$$

and similar equations for the components $(u_\downarrow(\mathbf{r},t), v_\uparrow(\mathbf{r},t))$. The single-particle Hamiltonian $h_{\uparrow,\downarrow}(\mathbf{r},t)$ is a partial differential operator (thus local), whose coefficients depend on all quasi-particle wave functions, and $\Delta(\mathbf{r},t)$ is the pairing field. The interactions with various applied external fields (spin, position and/or time-dependent) are described by including the appropriate potentials in the single-particle Hamiltonian.

We placed a UFG in its ground state in a cylindrical trap with periodic boundary conditions along the axis, almost flat inside and with soft lateral walls. We subsequently inserted various soft objects into the superfluid, and increased their impenetrability adiabatically while moving them through the fluid. In one set of theoretical experiments we sent a spherical projectile along the axis of the trap and generated quantized vortex rings [35]. In another set of experiments we introduced a rod, or two rods, or a rod and sphere and stirred the fluid by rotating the stirrers with constant angular frequency around the symmetry axis. In all cases the stirring occurred only for a finite time interval, after the which the stirrers were extracted adiabatically from the system and the UFG was left to evolve as a closed system. We varied both the stirring radius $R$ and the stirring angular frequency $\omega$ and thus controlled the speed $v_{stir} = R\omega$ of the stirrer(s). One can expect that if $v_{stir} \ll v_c$, where $v_c$ is the UFG critical velocity, the system will likely return to its initial state after the stirring is turned off. However, if $v_{stir} > v_c$ one expects that the whole system or a part of it will become normal and the superfluid order parameter will vanish in the corresponding regions of space. For a stirring with speeds $v_1 < v_{stir} < v_c$, where $v_1$ is some minimal stirring velocity, it is natural to expect that one or more vortices will be created. It is by studying such processes, and also the Higgs modes [13–15], that one can ascertain the qualitative differences between the predictions of traditional approaches and TDSLDA. An obvious limitation of TDSLDA is the neglect of the so called potential energy surface hopping, and an extension of this approach is warranted in order to accurately describe the long-time dynamics [34].

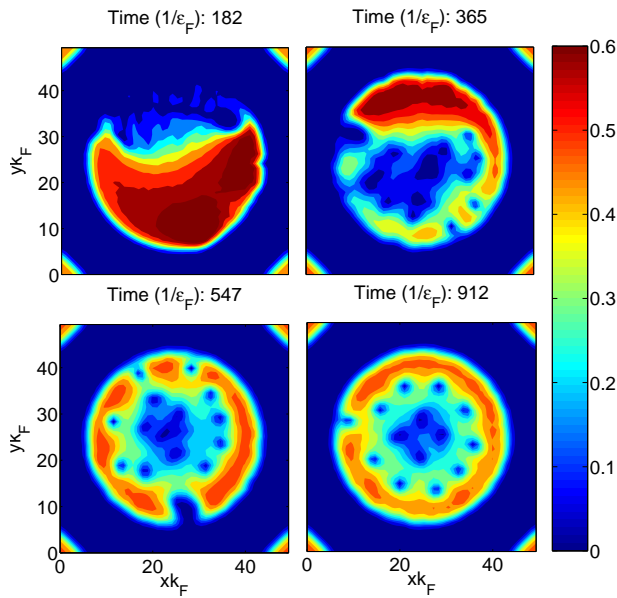

FIG. 1: The contour profiles of the pairing field $|\Delta(x, y, t)|$ of a unitary Fermi gas in a cylindrical container, stirred with a uniformly rotating rod, which is inserted and extracted adiabatically from the system. The gas, which has an almost uniform density distribution initially, is subsequently gathered almost entirely in front of the stirring rod. Both the number density and the pairing field are significantly depleted in the core of the vortices [36] and the number density profiles look very similar $|\Delta(x, y, t)|$. A plot (not shown here) of $\arg \Delta(x, y, t)$ reveals that the phase changes by $2\pi$ around a core as expected for quantized vortices.

While it was expected to generate a relatively small number of vortices when the stirring velocity is low, and that the number of vortices increases with the stirring velocity, a great number of features of the dynamic vortex generation are unexpected. The fact that UFG is highly compressible results in very large time-dependent variations of the number density. Often the entire mass of the system is gathered in front of the stirrer leaving little matter behind it even at subsonic speeds. The gas occupies less than half of the available volume, even when the excluded volume by the stirrer is quite small and it comes as a surprise that the system does not lose quantum coherence under such a violent perturbation. Moreover, the system organizes itself in an almost perfect vortex lattice after the stirring is turned off. Even more surprising is that the system remains superfluid even when stirred at supercritical speeds. We have observed that the system forms a vortex lattice even if stirred with speeds up to $v_{stir} = 0.65v_F > v_c \approx 0.365v_F$, see [35]. In a gas, unlike a liquid which has a rather well defined density, the increase of the density of the cloud during the stirring process leads to an increased local critical velocity and thus to a stable superflow at speeds larger than the nominal Landau's critical velocity.

Fig. 2 illustrates the generation of a very elusive phenomenon, the formation of vortex rings by a spherical projectile flying with velocity $0.2v_F < v_c$ along the axis of a cylindrical trap. The two vortex rings are formed with slightly different radii and thus they propagate with different drift velocities. The smaller vortex ring is formed first and catches the larger vortex rings. When the rings separate, they emerge with slightly different radii. In Fig. 3 we demonstrate the microscopic dynamics of the reconnection of two vortex lines, which may be at the origin of quantum turbulence [5, 11]. In this case two vortex lines cross at two points, which results in the vortex lines exchanging finite segments. In essentially all the cases we studied, regular sound waves (phonons) were also excited. In a separate simulation of an off-center sphere passing through the system [35] we have observed that the recombination of two vortex lines is sometimes a two step process, two lines cross and than re-cross before separating. In addition, when we generated vortices using a rod and a sphere [35], the translational symmetry was destroyed and one could clearly observe the excitation of the Kelvin modes, waves running along the vortex lines. With time the Kelvin modes apparently transfer their energy to phonons from short length scales to larger length scales - a mechanism which is still controversial in the quantum turbulence community [11]. However, this particular aspect requires further and more detailed analysis. In the cases we studied, this energy transfer might have been partially due to the imposition of periodic boundary conditions along the third axis.

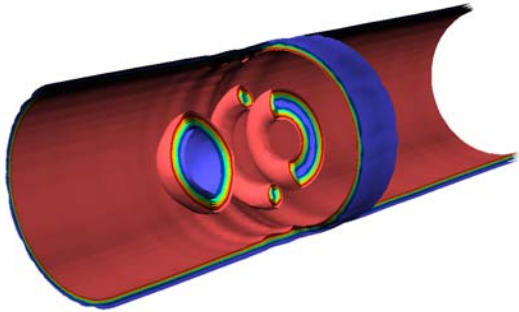

FIG. 2: A spherical projectile flying along the symmetry axis leaves in its wake two vortex rings.

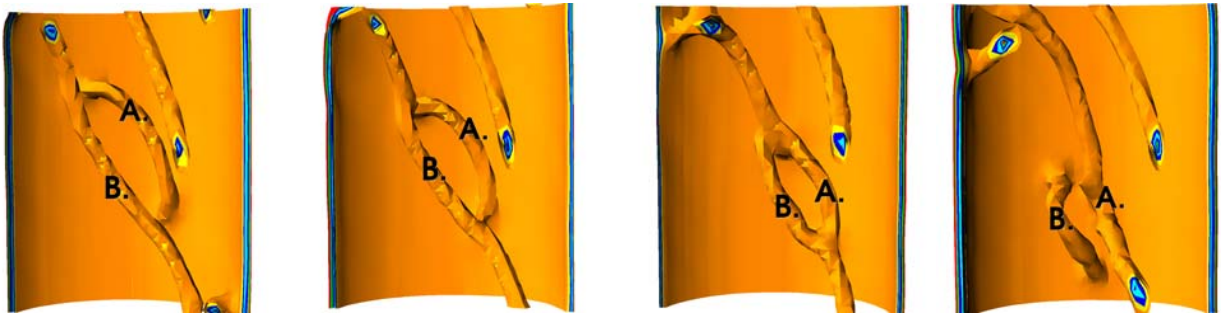

FIG. 3: Two vortex lines approach each other, connect at two points, form a ring and exchange between them a portion of the vortex line and subsequently separate. The segment A which initially belonged to the vortex line attached to the wall is transferred to the long vortex line after reconnection and vice versa.

The various excitation scenarios illustrated here for a UFG demonstrate the power of this new framework, the richness of the phenomena which are still awaiting to be fully explored and the extreme flexibility of this new approach. Due to the complexity of the full 3D TDSLDA equations these phenomena have never been addressed even in the meanfield approach. Within the theoretically superior TDSLDA framework one can address with high accuracy on a microscopic basis the UFG superfluid dynamics, and for the first time contemplate a study of the quantum turbulence at zero temperature, where dissipative processes are greatly inhibited.

**Acknowledgment** We thank G.F. Bertsch for discussions. Support is acknowledged from the DOE grants DE-FG02-97ER41014 and DE-FC02-07ER41457, the contract No. N N202 128439 from Polish Ministry of Science, and NKBRSF of PR China under Grant No. 2011CB921503. Calculations have been performed on UW Athena cluster, Hyak UW cluster funded by the NSF MRI grant PHY-0922770, Franklin supercomputer (Cray XT4, US DOE NERSC, grant B-AC02-05CH11231), and JaguarPF supercomputer (Cray XT5, US DOE NCCS, contract DE-AC05-00OR22725).

Supplemental online material for

*Generation and Dynamics of Quantized Vortices*

*in a Unitary Fermi Superfluid*

*Aurel Bulgac, Yuan-Lung (Alan) Luo, Piotr Magierski,*

*Kenneth J. Roche, and Yongle Yu*

Many technical details and the relations to previous works of other authors have been described in significant detail in the review:

A. Bulgac, M.M.Forbes, and P. Magierski, *The Unitary Fermi Gas: From Monte Carlo to Density Functionals*, arXiv:1008.3933, to appear in *BCS-BEC Crossover and the Unitary Fermi Gas* (Lecture Notes in Physics), edited by W. Zwerger (Springer, 2011) and we refer the interested reader to consult this source for additional references as well.

This supplement consists of three sections:

1) A brief description of 27 movies with various perspectives of the 3D simulations of a UFG excited by stirring with various external probes in several geometries. We show in real time the generation of quantized vortices and the formation of vortex lattices in restricted geometries, the generation of vortex rings and their interaction with one another and the trapping potential, the recombination of vortex line, the destruction of the superfluidity by energetic stirring and related processes.
2) A brief description of the energy density functional and the time-dependent formalism.
3) Outline of the numerical methods and a brief description of the parallel implementation on the NCCS Cray supercomputer JaguarPF.

# Brief description of various movies of the UFG simulations

There are two main types of simulations that we have performed. In the first batch of constrained 3D dynamics we have enforced the homogeneity of the system in the z-direction. In the second batch we have allowed for a fully unrestricted 3D dynamics. In both cases we have placed a relatively large number of particles in a cylindrical trap that is axially symmetric with respect to the z-axis. The stirrers used in the first set of simulations did not break the homogeneity in the z-direction and only disturbed the system in the xy-plane. In the second set of simulations we have allowed the stirrers to break the homogeneity in the z-direction as well.

Please notice that hyperlinks to various simulations are clearly marked and active throughout the entire text. In particular, at http://www.phys.washington.edu/groups/qmbnt/UFG one can access the full set of our simulations. If the viewer encounters difficulties playing these simulations we suggest trying the VLC player for a unix OS, Windows Media Player or sometimes Quick Player as well for a Windows or MAC computers. In various tests we performed we had the best experience with Firefox and Windows Media Player. IE appears to perform rather well too. For the best experience we suggest to open the present document in a separate window and open various movies either from the html table http://www.phys.washington.edu/groups/qmbnt/UFG/nt-ufg.html or from the actual directory http://www.phys.washington.edu/groups/qmbnt/UFG/ In order to avoid long download times and for faster playback times and effective display of the player controls, we strongly suggest to download the ***nt-ufg-dat.tar.gz*** file with these movies, in which one can find also the nt-ufg.html table with the embedded links to them.

## *Constrained 3D Simulations (homogeneity in z-direction)*

The stirrer in the first type of simulations was a repulsive cylindrical potential (which experimentally can be generated with a blue detuned laser beam) of Gaussian shape of full width 2 (in lattice units), placed at a radius $R = 6, 10$ or $15$ as specified in the legend in the next two figures or in the names of the corresponding movies. The stirrer moved with constant angular velocity $\omega$ and the resulting linear velocity of the stirrer was $v_{stir} = \omega R$ which is specified in the name of each movie. For example the movie N32P300_r6_0.4_V9 is for a system of 300 particles, in a simulation box $32^3$ with a unit lattice constant and a stirring radius $R = 6,$ a linear stirring velocity of $v_{stir} = \omega R = 0.4 v_F$, and in which we observe the formation of 9 vortices. In each of these movies there are four panels. The upper left panel displays the total instantaneous potential experienced by the fermions; the upper right panel displays the number density distribution. The lower left panel shows the magnitude of the pairing field, while the lower right panel shows the phase of the pairing field. On top of each frame we sow the product of the time with Fermi energy $t\varepsilon_F$. When vortices are formed one observes a significant number density and pairing field depletion in the core of the vortex and also that the phase of the pairing field changes by $2\pi$ when one goes around the center of each vortex. As images speak for themselves we shall not describe in detail these movies and only summarize several common aspects. The controls allows the viewer to either let the movie run at its speed, stop it, or move the frames both forward and backward in time. One can clearly see when the stirrer is turned on and off during the stirring process and its effect on the system. In all cases the central density of the cloud corresponds to a local Fermi momentum $k_F \approx 1$ in lattice units $(\hbar = m = 1)$. The height of the stirrer potential energy was chosen equal to $\varepsilon_F = k_F^2 / 2$, and the local number density at the center of the cloud is

$$n = \frac{k_F^3}{3\pi^2}.$$

A summary of the first kind of simulations is presented in the following two figures.

Figure 1 shows the excitation energy in various simulations as a function of the linear stirring velocity. On one hand, these systems remain superfluid while developing vortex lattices when stirred with velocities significantly exceeding the nominal Landau critical velocity, which for a UFG is $v_c = 0.365 v_F$. On the other hand, if the stirring velocity is small, then the system essentially returns to its initial state after the stirrer is taken out.

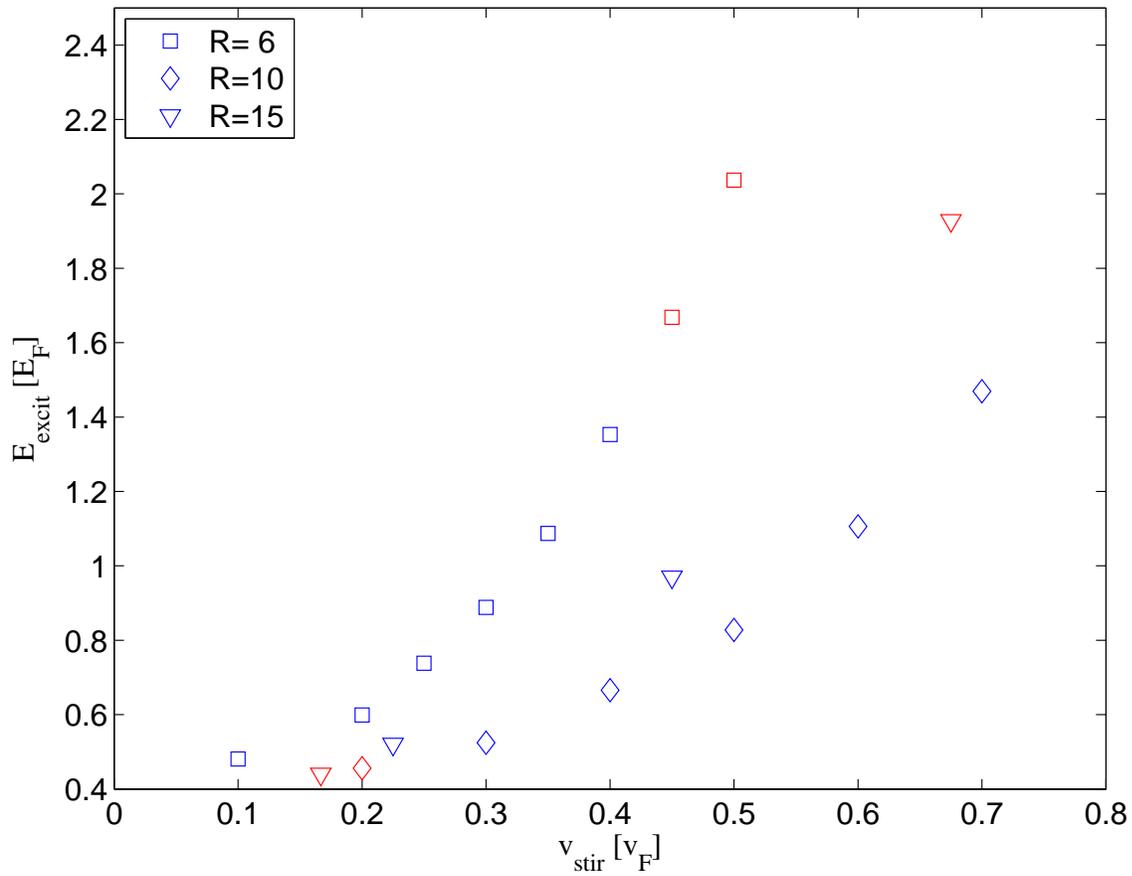

Figure 1. Here we show a summary of the vortex generation simulations of UFG confined to a cylinder with periodic boundary conditions along the symmetry axis. With squares we display the

*results obtained in a simulation box $32^3$, when the center of the stirrer was at radius $R = 6$, with diamonds the case $R = 10$, and with triangles the simulation in the $48^3$ box with $R = 15$. The units for the energy of the system are $E_F = 0.6\varepsilon_F N$, where $N$ is the total particle number in the trap. The blue symbols are cases when vortices have been created. With red symbols are cases when no vortices were generated, either because the stirring velocity was too small or too large.*

We can compare the excitation energy in the presence of $N_v$ vortices to the rigid body rotation energy with the same total angular momentum

$$E_{rigid} = \frac{L^2}{2I} \propto N^{1/3} N_v^2 ,$$

where $L_z = \frac{N}{2} N_v$ is the total angular momentum of $N_v$ vortices and $I \propto N^{5/3}$ is the rigid body moment of inertia of the system. The moment of inertia was calculated using the ground state number density. During the stirring, especially when a relatively large number of vortices are created, the density of the UFG at the center is depleted and the actual rigid body moment of inertia is larger than the value calculated with the ground state number density. The actual excitation energy for a given number of vortices is in each case larger, which we attribute in part to the fact that additional phonon modes are excited as well, among them vibrations of the vortex lattice. When we divide the excitation energy by $N^{1/3}$ the particle dependence on the total particle number dependence is removed and all energies lie almost exactly on the same parabola as a function of the total number of vortices for all systems.

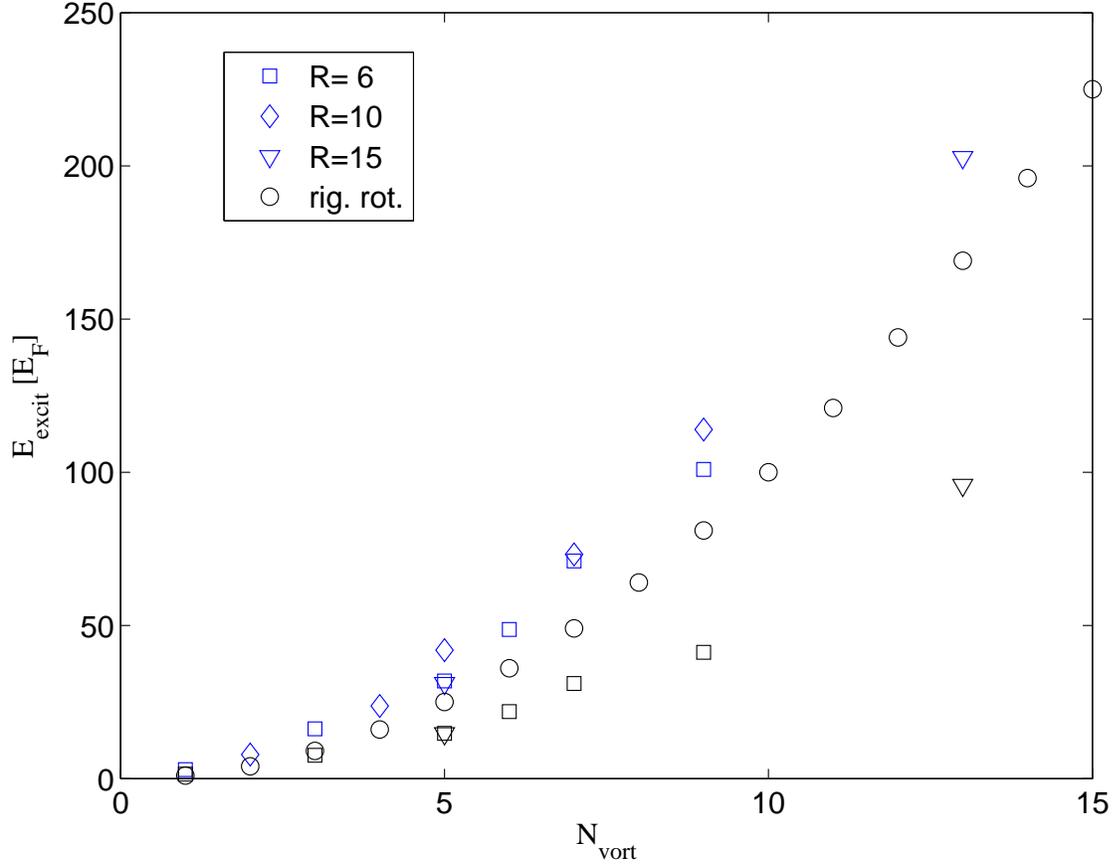

Figure 2. The excitation energy of the UFG scaled with the particle number $N^{1/3}$ and the units for the energy are chosen so as to remove the particle dependence $E_F = 0.6\varepsilon_F N^{1/3}$.

If we further calculate the collective kinetic energy at the end of the stirring process we obtain the values shown with black squares and triangles in Figure 2. We have calculated the collective kinetic energy using the following equations

$$E_{\text{coll. kin.}} = \int d^3 r \frac{\vec{j}^2(\vec{r},t)}{2n(\vec{r},t)},$$

$$\vec{j}(\vec{r},t) = \text{Im}\left[i \sum_k v_k^*(\vec{r},t)\vec{\nabla} v_k(\vec{r},t)\right].$$

We observe that the collective kinetic energy is practically constant after the stirring is performed only when the number of vortices is $\leq 5$ for the duration of our simulations. For a higher number of vortices this energy is still evolving with time and decreasing slowly after the stirrer is turned off, as these systems have not reached yet a steady state, given our relatively short simulation times. The collective kinetic energies are noticeably smaller than both the rigid body kinetic energies (since the actual moment of inertia is increased in a system in rotation with respect to the ground state value this is expected) and even smaller than the actual excitation energies. We conclude that the collective kinetic energy is a relatively small part of the excitation energy of a system with vortices. The reminder of the excitation energy is thus due to the strong number density redistribution for a system with many vortices and it is mostly due to the vortex-vortex interaction. It seems that the vortex-vortex interaction has two effects. One effect is the strong pair wise interaction among vortices, and this fact is reflected in the essentially quadratic behavior of the excitation energy with the number of vortices. The other effect consists in the significant depletion of the number density in the region where the vortex cluster is formed and the increase of the number density near the wall of the trap.

In the case of large number of vortices one can notice that the pairing field in the presence of vortices does not reach the same maximum values as in the ground state. One might naively expect that the pairing field could attain values larger than in the ground state, since the number density close to the walls is actually larger than the central number density in the ground state. For example, in the simulation [N32P300_r10_0.7_V9](N32P300_r10_0.7_V9) the vortices are rather faint, though rather prominent on the lower left panel for the phase of the pairing field. And even though the number density at the edges of the system is clearly higher than the central number density in the ground state, the values of the pairing field are noticeably lower. One might argue whether in this case one really has 9 vortices or whether the superfluid phase has actually disappeared in the center of the system. The comparison of this simulation with the somewhat crisper one [N32P300_r10_0.6_V7](N32P300_r10_0.6_V7) however lends support to our conclusions that 9 vortices have been formed in this case. It thus becomes clear that the pairing field outside the vortex cores in fermion systems in the presence of a large number of vortices is qualitatively different from the pairing field in the ground state.

At relatively low stirring velocities, see [N32P300_r10_0.2_V0](N32P300_r10_0.2_V0) and [N48P1800_r15_0.1667_V0](N48P1800_r15_0.1667_V0), vortices are not generated. Apparently this behavior depends on the size of the trap, of the stirrer and of its position.

On a different note, as the simulation [N48P1800_r15_0.4500_V13](N48P1800_r15_0.4500_V13) shows, superfluidity clearly survives at driving velocities well above the nominal Landau critical velocity. We will see a related example among the full 3D simulations discussed below. This qualitatively new aspect is due to the high compressibility of these systems, which is unlike the behavior of electrons in condensed matter or of liquid helium, whose density outside the vortex cores is left largely unchanged in the presence of vortices. The ability of the present formalism to correctly describe the transition of a superfluid system to a normal state is a qualitatively new feature of the present formalism, which is not within the range of applicability of the traditional phenomenological approaches. While Ginzburg-Landau and Gross-Pitaevskii type of approaches could in principle describe the generation of quantized vortices, Landau two-fluid hydrodynamics will clearly fail, as there is no Planck's constant in this approach. Landau's critical velocity is also absent in the two-fluid hydrodynamics approach and thus the transition from superfluid to normal phase is not described correctly.

In the last five movies of this series we demonstrated the role of geometry in the generation and dynamics of vortices. We have stirred a UFG in an elliptic cylindrical trap and observed that stable quantized vortices fail to form at subcritical velocities already, but not if the stirring velocity is sufficiently small. We have also observed that if the stirrer is performing off-centered circular motion vortices are formed in a circular cylindrical trap.

## *Full 3D simulations*

In the full 3D simulations we have released the constraint that the system remains homogeneous in the z-direction and we could thus study for the first time in literature the full unrestricted dynamics of a superfluid fermionic system. We stirred the ground state systems which described above, with various external time-dependent potentials and we observed a wide variety of excitations. In all these movies we display isosurfaces of the magnitude of the pairing field and in the frames we show the time in units of $1/\varepsilon_F$.

[**nt-ball-rod-dns**](#), [**nt-ball-rod-dns-pln**](#), [**nt-ball-rod-thin-angl**](#) (simulation box 32x32x32)

In all these movies we show the dynamics of the magnitude of the pairing field when the system is stirred with a rod parallel to the $z$-axis and a sphere diametrically opposite to it. Five vortices are created by the end of the stirring. Because of the presence of the sphere, the translational symmetry along the $z$-axis is broken and one can clearly see the excitation of the Kelvin modes (waves along the vortex lines). It is notable that vortices tend to cling to the sphere, which is likely the mechanism of the Kelvin mode excitation. The last two vortices are created at almost the same time and they stick to each other for quite some time before separating. With time the amplitude of the Kelvin modes decays and the vortices tend to become almost straight and the energy of the Kelvin modes is transferred to phonons. One can conclude that energy is transferred from higher to lower wave vectors. It is likely that this tendency is somewhat enhanced in this system, due to the presence of the periodic boundary conditions along the $z$-axis, which thus favors straighter vortex lines. Clearly no equilibration of the vortex lattice is achieved during the relatively short time of the simulation.

[nt-ball-c](#), [nt-ball-c-w](#) (simulation box 32x32x96)

There are two vortex rings formed by a spherical projectile moving with a velocity $0.2 \text{v}_F$. The first vortex ring is smaller and thus is moving faster than the second vortex ring. The small ring catches up with the slower vortex ring and during their interaction their radii oscillate and the smaller ring eventually passes through the larger one. The initial and final sizes of these two rings differ.

[nt-ball-longr](#) [nt-ball-longr-w](#) (simulation box 32x32x196)

The spherical projectile with velocity $0.2v_F$ is launched exactly along the axis of the cylindrical trap and it generates at least four rather stable vortex rings of slightly different radii, and thus speeds. One can observe the formation of a very small vortex ring, which propagates as expected rather fast and while if passes through the larger ring is essentially ``eaten up'' by the larger vortex ring. Apart from vortex rings one can also observe the generation of phonons, which propagates with the velocity of sound.

[nt-small-ball](#) (simulation box 32x32x96)

A smaller spherical projectile is sent along the axis of the cylinder with velocity $0.2v_F$. Such a projectile would have the tendency to generate vortex rings of smaller radii, but since such vortex rings would have the tendency to move faster than their source, they do not survive.

[nt-big-ball](#), [nt-big-ball-w](#), [nt-big-ball-angle-w](#) (simulation box 32x32x96)

In this simulation we sent a spherical projectile with velocity $0.2v_F$ and radius very close to the trap radius, but a bit smaller. One observes the excitation of vortex rings and phonons of quite significant amplitudes and their nonlinear couplings. The conditions are not optimal for the formation of stable vortex rings, since in the case of such a big sphere the vortex ring would tend to be slow and of a size likely exceeding significantly the radius of the trap.

[nt-ssonic](#) (simulation box 32x32x96)

A spherical projectile is launched exactly along the axis of the trap with velocity $0.45v_F > v_c$. The system remains superfluid at all times in spite of the fact that the Landau criterion is violated.

[nt-ball-oc-v1](#), [nt-ball-occ-v2](#), [nt-ball-oc-wz](#), [nt-ball-oc-bot](#), [nt-ball-oc-top](#), (simulation box 32x32x96)

In this simulation a spherical projectile is launched along the axis of a cylinder with $v_{sph} = 0.2v_F < v_c$. Various movies show different perspective of the same simulation, when a spherical projectile was launched slightly off center but parallel to the axis of symmetry of the trap. One can observe the formation of several vortex lines and vortex rings. The vortex lines are attached to the wall of the trap, the vortex rings are slight off center, are not planar and at an angle with the cylinder axis. Various vortex lines and rings propagate at different velocities and at various times they cross and one can clearly see vortex reconnections. We show a vortex recombination in several close-ups and from a couple of different perspectives. These are the first microscopic simulation of the vortex reconnection in a Fermi superfluid. This process was conjectured by Feynman to be the source of quantum turbulence. At very low temperatures dissipative processes are strongly suppressed and the classical mechanism of turbulence is thus not possible

[nt-collide](nt-collide) (simulation box 32x32x96)

In this simulation we have sent simultaneously two spheres, both slightly off center towards each other. One can observe the generation of both vortex rings and vortex lines attached to the walls, as well as several vortex reconnections.

[nt-twst](nt-twst), [nt-twst-w](nt-twst-w), [nt-twst-wz](nt-twst-wz), [nt-twst-msh](nt-twst-msh),  (simulation box 32x32x96)

We have used a helically twisted rod to stir the UFG. After a couple of turns several vortex helical lines are generated. Since the system resides on a spatial lattice with periodic boundary conditions in both $x-$ and $y-$directions the vortex lines closer to the wall experience some small perturbations, in spite of the thickness of the potential barrier between traps in neighboring cells, and these vortex lines start sticking to the walls. One can observe the emergence of quite numerous rather complicated vortex reconnections, while at the same time other vortex lines show the tendency towards becoming straighter. Some of these movies show close-ups of when a whole segment is exchanged between two vortex lines, a process more complex than the usual vortex recombination conjectured to occur in superfluids.

# The energy density functional and the time-dependent formalism

In condensed matter and chemistry calculations the leading method is the Density Functional Theory (DFT). In the DFT method one replaces the solution of the Schrödinger equation for an N-electron system requiring the solution of a partial differential equation in a 3N-dimensional space with a system of N nonlinear coupled 3D-equations. This simplification is achieved by introducing an energy density functional, the variation of which provides the energy and the electron density spatial distribution of the ground state of the system. The existence of the energy functional was proven by Kohn, Hohenberg and Sham in 1964-1965 and this achievement was recognized with the Nobel Prize in chemistry to W. Kohn in 1998. Further theoretical developments and the extension of the DFT formalism have led today to the study of the properties of a large number of excited states of electron systems within the framework of the Time-Dependent-DFT (TDDFT) formalism. This highly successful theoretical formalism is limited to so-called normal systems, however, and is essentially impossible to apply to superconductors in particular. In the early 1980s, a nonlocal extension of the DFT for the study of superconductors was suggested, and has been implemented during the last few years by E.K.U. Gross (Berlin) and his group for the study of a few systems.

The great success of DFT applied to normal electronic systems was ensured by the introduction of LDA (Local Density Approximation) form of DFT by Kohn and Sham, which leads to <u>local</u> nonlinear coupled partial differential equations, as opposed to <u>nonlocal</u> nonlinear coupled integral-partial differential equations. This great advantage of DFT was nullified in the extension of the DFT formalism implemented by Gross and collaborators. We have developed instead a local extension of DFT to superfluid systems, the Superfluid LDA (SLDA) and its generalization to time-dependent phenomena (TDSLDA).

The SLDA is based on the following energy density functional for the ground state for an unpolarized unitary Fermi gas:

$$E_{gs} = \int d^3r \left[ \frac{\hbar^2}{2m} \tau(\vec{r}) + g_{eff}(n(\vec{r})) |\kappa(\vec{r})|^2 + \varepsilon(n(\vec{r})) + V_{ext}(\vec{r}) n(\vec{r}) \right],$$

$$n(\vec{r}) = 2\sum_n |v_n(\vec{r})|^2, \quad \tau(\vec{r}) = 2\sum_n |\vec{\nabla} v_n(\vec{r})|^2, \quad \kappa(\vec{r}) = \sum_n v_n^*(\vec{r}) u_n(\vec{r}),$$

where $n(\vec{r})$ is the number (normal) density, $\tau(\vec{r})$ is the kinetic energy density, and $\kappa(\vec{r})$ is the anomalous density. The anomalous density is proportional to the superfluid order parameter and vanishes in the normal phase. These densities depend on the quasiparticle wave functions $(u_n(\vec{r}), v_n(\vec{r}))$. $V_{ext}(\vec{r})$ is an arbitrary external potential in which the system might reside, $\varepsilon(n(\vec{r}))$ is the normal part of the interaction energy density and $g_{eff}(\vec{r})$ is the renormalized coupling strength of the pairing correlations. Varying the quasiparticle wave functions we obtain the Bogoliubov-de Gennes-like equations (BdG)

$$\begin{pmatrix} -\dfrac{\hbar^2 \Delta}{2m} + U(n(\vec{r})) + V_{ext}(\vec{r}) - \mu & \Delta(\vec{r}) \\ \Delta^*(\vec{r}) & \dfrac{\hbar^2 \Delta}{2m} - U(n(\vec{r})) - V_{ext}(\vec{r}) + \mu \end{pmatrix} \begin{pmatrix} u_n(\vec{r}) \\ v_n(\vec{r}) \end{pmatrix} = E_n \begin{pmatrix} u_n(\vec{r}) \\ v_n(\vec{r}) \end{pmatrix},$$

which must be solved self-consistently in order to determine the ground state properties of a given system. These equations represent an infinite set of nonlinear coupled partial differential eigenvalue equations. Upon discretization, these equations become nonlinear coupled matrix equations that are solved iteratively. The convergence rate can be accelerated by various techniques, such as Broyden's line search algorithm.

The energy density functional in dilute Fermi gas systems has been determined with high accuracy in Quantum Monte Carlo studies, and the corresponding Equation of State has also been confirmed experimentally. Thus, the essential elements to study the time-dependent problem are available. The derivation of the energy functional includes three free parameters. By requiring the (homogeneous) solution to the BdG equations match Monte Carlo calculations (energy, pairing field, and quasiparticle spectrum) these three parameters are determined. The great accuracy of the SLDA description is illustrated in Figure 3 (for details see the review cited at the beginning) and it is limited so far by the inaccuracies of the Monte Carlo calculations.

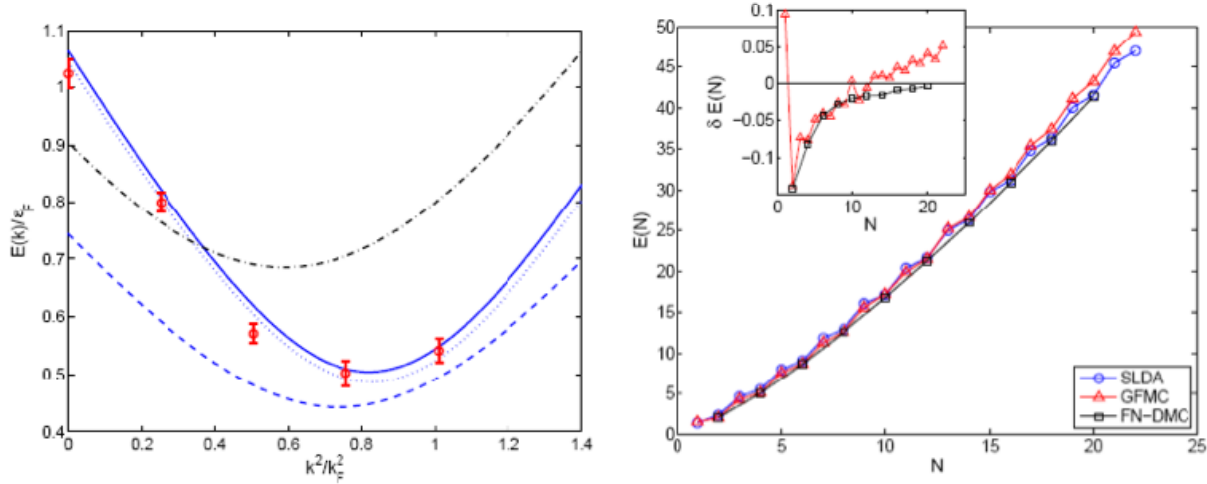

*Figure 3. (left panel) SLDA (solid blue) compared to Monte Carlo calculation (red + errorbars).*

*(right panel) SLDA energies versus Quantum Monte Carlo results for systems with various even and odd particle number in a harmonic trap.*

In the time-dependent case we consider the variation of a different functional,

$$L = \int dt\, d^3r \left[ i\hbar \sum_n v_n(\vec{r},t) \frac{\partial v_n^*(\vec{r},t)}{\partial t} \right]$$

$$- \int dt\, d^3r \left[ \frac{\hbar^2}{2m} \tau(\vec{r},t) + g_{eff}(n(\vec{r},t))|\kappa(\vec{r},t)|^2 + \varepsilon(n(\vec{r},t)) + V_{ext}(\vec{r},t) n(\vec{r},t) \right],$$

$$n(\vec{r},t) = 2\sum_n |v_n(\vec{r},t)|^2, \quad \tau(\vec{r},t) = 2\sum_n |\vec{\nabla} v_n(\vec{r},t)|^2, \quad \kappa(\vec{r},t) = \sum_n v_n^*(\vec{r},t) u_n(\vec{r},t).$$

All quantities are functions of time. In general the structure of this functional is more complicated, and Galilean invariance requires that we use a slightly different form for the kinetic energy density, which we will not discuss here and in which case an effective mass is present. For the sake of simplicity we omit the equations of motion for the quasiparticle wave functions in their most general form (when an effective mass different from the bare of mass is present) and we refer the interested reader to the above mentioned review. The time-dependent evolution equations for the quasiparticle wave functions $(u_n(\vec{r},t), v_n(\vec{r},t))$ become

$$i\hbar \frac{\partial}{\partial t} \begin{pmatrix} u_n(\vec{r},t) \\ v_n(\vec{r},t) \end{pmatrix} = \begin{pmatrix} -\frac{\hbar^2 \Delta}{2m} + U(n(\vec{r},t)) + V_{ext}(\vec{r},t) - \mu & \Delta(\vec{r},t) \\ \Delta^*(\vec{r},t) & \frac{\hbar^2 \Delta}{2m} - U(n(\vec{r},t)) - V_{ext}(\vec{r},t) + \mu \end{pmatrix} \begin{pmatrix} u_n(\vec{r},t) \\ v_n(\vec{r},t) \end{pmatrix}.$$

These represent an infinite set of nonlinear coupled partial differential equations of evolution.

In all simulations the system started in the ground state and various objects were introduced into the system by varying the strength of their repulsive interaction adiabatically as illustrated in Figure 4 below.

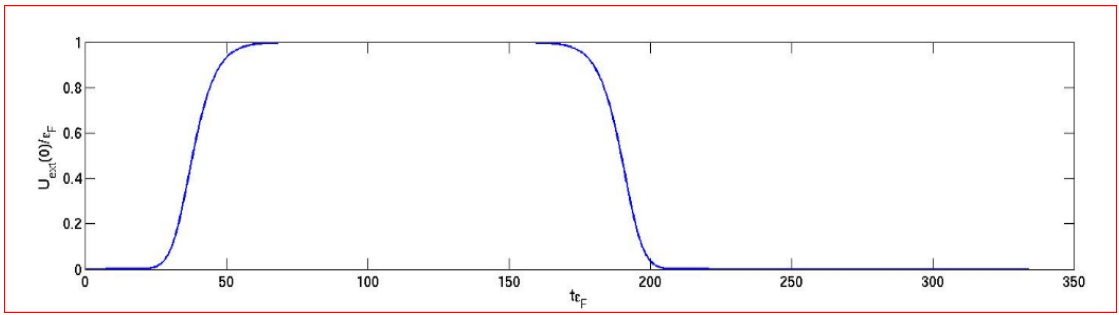

Figure 4. The strength of the external stirrer as a function of time in a typical simulation.

The length of the stirring time interval and the maximum strength of the external potential varied from simulation to simulation and can be visualized by observing specific movies discussed below and accessible online. During the stirring period energy is pumped into the UFG. The energy of the system remains constant after the stirring is turned off, as expected for an isolated system, see plot below for a typical behavior of the total energy of the system as a function of time.

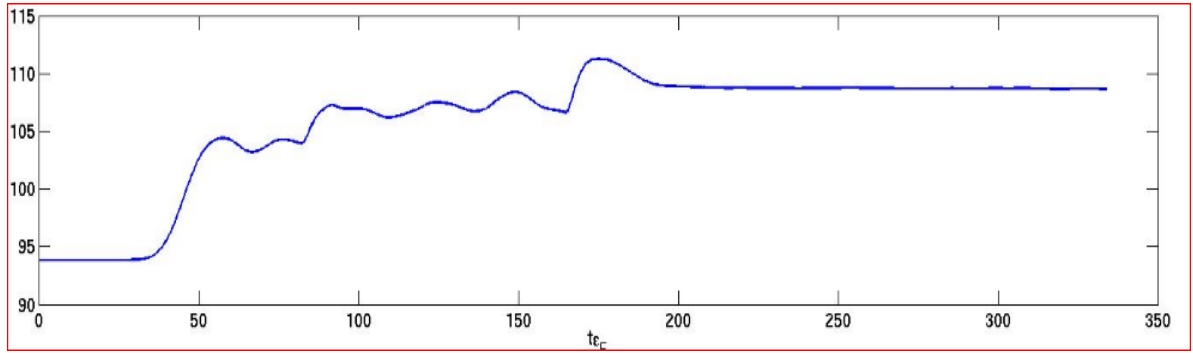

*Figure 5. The total energy of 300 particles as a function of time in a typical stirring simulation.*

The following figure show several frames of a simulation, in which a UFG in a cylindrical trap was excited by a rod parallel to the cylinder axis and a sphere diametrically opposite to it, both rotating with the same angular frequency.

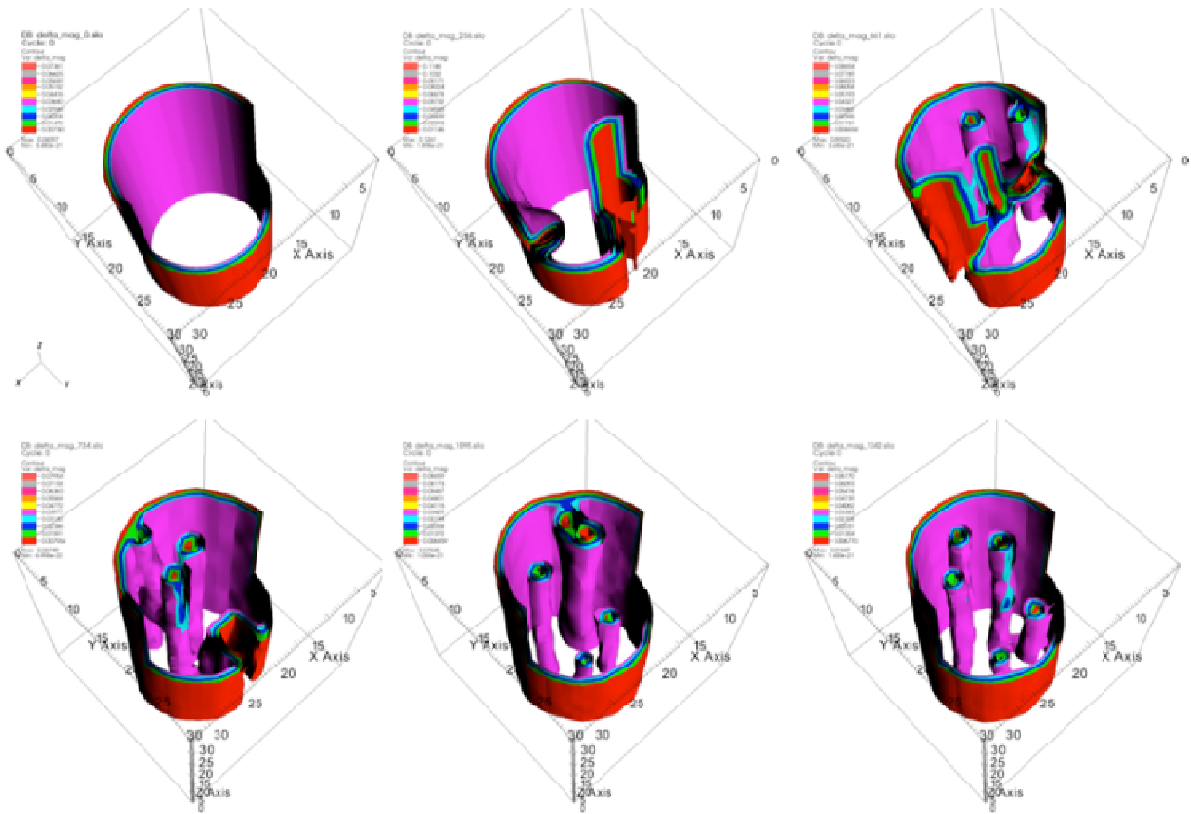

*Figure 6. Vortex formation and dynamics of a UFG system of 300 particles in an external cylindrical trap. The gas is excited through stirring with a ball and rod system and left to evolve and eventually five vortices clearly form.*

# Numerical methods and a brief description of the parallel implementation on the NCCS Cray supercomputer JaguarPF

These simulations were performed in boxes with spatial lattices 32x32x32, 32x32X96, 32x32x192, 48x48x48. The number of quasiparticle wavefunctions $\left(u_n(\vec{r},t), v_n(\vec{r},t)\right)$ was approximately 9,500, 35,000, 70,000 and 41,000 respectively. We have integrated the TDSLDA equations for hundreds of thousands of time steps and even million of time steps sometimes, depending on the simulation. The times are reported in units of the inverse Fermi energy. The trapping potential was chosen as

$$U_{trap}(x, y, z) = U_0 \left[ \sin\left( \frac{\sqrt{x^2 + y^2}}{r_0} \right) \right]^n$$

with various rather large values of the exponent $n$ and periodic boundary conditions were imposed along the $z$-axis. The UFG central density was chosen such that the local value of the Fermi wave vector at the center of the trap was $k_F \approx 1$ in units corresponding to a lattice constant $a = 1$. The shape of the stirrers varied from one or two rods parallel to the $z$-axis, to a rod and a sphere, or simply a sphere, and sometimes a twisted helical rod.

Our problem is a system of coupled complex time-dependent nonlinear partial differential 3D equations with periodic boundary conditions. The solutions are 2-component quasiparticle wavefunctions that are represented on a discrete three-dimensional spatial lattice with a lattice constant a, N lattice points in each spatial direction (although the spatial dimensions of the lattice do not in general need to be identical), and periodic boundary conditions. In the static (solver) SLDA code we calculate first- and second-order spatial derivatives using a matrix representation of the FFT (Fast Fourier Transform) according to the formulas given below, where F(x) is the interpolating function of the DVR (Discrete Variable Representation) method. Using this interpolating function F(x) we evaluate the corresponding matrix representation of the first and second derivatives, as shown in the following equations:

$$F(x) = \sum_{l=0}^{N-1} \frac{1}{N} \exp\left( \frac{ixk_l}{Na} \right) = \frac{1}{N} \frac{\sin\left(\frac{\pi x}{a}\right) \exp\left(-i \frac{\pi x}{Na}\right)}{\sin\left(\frac{\pi x}{Na}\right)}$$

$$k_l = -\frac{\pi}{a} + \frac{2\pi l}{Na}, \quad l = 0,...,N-1$$

$$x_n = na, \quad n = 0,...,N-1$$

$$F(x_n - x_m) = \delta_{nm}$$

$$\nabla_{nm} = F'(x_n - x_m) = \frac{\pi}{Na}(-1)^{n-m} \left[ (1-\delta_{nm})\cot\left(\frac{\pi(n-m)}{N}\right) - \frac{i}{N} \right]$$

$$\Delta_{nm} = (\nabla^2)_{nm} = F''(x_n - x_m) = \frac{\pi^2}{2N^2 a^2} \frac{(-1)^{n-m}(1-\delta_{nm})}{\sin^2\left(\frac{\pi(n-m)}{N}\right)} - \frac{\pi^2}{3a^2}\left(1 + \frac{2}{N^2}\right)\delta_{nm}$$

For lack of a better notation, we have used above the symbols usually reserved for the gradient and Laplacian in 3D for the matrix representation of the first and second derivative operators in 1D. The eigenvalues of the above discrete matrices $\nabla$ and $\Delta$ are exactly $ik_j$ and $-k_j^2$, as expected. We need to determine the full spectrum of the 3D Schrödinger equation. In each self-consistent step we perform the diagonalization of a Hermitian matrix of size $2N^3 \times 2N^3$. We implemented multi-variable optimization schemes based upon imaginary time evolution of the system with linear (modified Broyden) mixing.

For systems that are homogeneous in one spatial direction, say the z-component, the quasiparticle wave functions have a simpler structure,

$$\begin{pmatrix} u_n(x,y,t) \exp(ik_n z) \\ v_n(x,y,t) \exp(ik_n z) \end{pmatrix},$$

and densities and potentials do not depend on the z-variable. Experimentally, this situation can be approximately realized in very elongated cigar shaped traps for example.

The generalization of these formulas to the 3D case and coupled nonlinear equations of many wave functions with several components is straightforward. In the DVR method only the kinetic energy is represented as a matrix, while all local potentials appear as diagonal matrices, thus making the evaluation of all these quantum operators simple to implement numerically. We have implemented solver(s) that exploit such homogeneities in the z-component of the problem geometry and can execute each $k_z$ value independently (there are $N_z/2 + 1$ such values in general) within a self-consistent iteration. This amounts to forming independent parallel work groups, each of which simultaneously numerically diagonalizes a reduced dimension (i.e., $n = 2N_xN_y$ versus $2N_xN_yN_z$ in the unitary case) matrix each self-consistent iteration. It should be clear that one achieves a nearly perfect strong scaling curve when computing $O(N_z)$ diagonalizations simultaneously versus in sequence. Furthermore, the cost of a single diagonalization is reduced by $O(N_z^3)$ over the general problem. We were originally utilizing a parallel QR-based subroutine to achieve the diagonalizations within each self-consistent iteration, but this was not performing well at scale. However, there exists a much more efficient method due to Cuppen that applies a divide-and-conquer procedure to achieve more precise numerics at nearly four times the speed on all problems we have tested on the Cray XT5 supercomputer (JaguarPF) at the US Department of Energy National Center for Computational Sciences at Oak Ridge National Laboratory. The method reduces the original problem to two independent symmetric tri-diagonal eigenvalue problems of dimension k and N – k. This reduction can be repeated until a stopping condition is satisfied, at which point QR iteration is applied on a much smaller system.

| *Nx(=Ny=Nz)* | *#Diags / Group* | *T[s] Converge* | *Speed-up* | *PEs / Group* |
|---|---|---|---|---|
| 86 | 2 | 121.25 | | 256 |
| 86 | 1 | 57.22 | 2.118 | |

*Figure 7. Strong scaling over self-consistent iteration for systems with homogeneity in one coordinate.*

In TD-SLDA we evaluate only the action of the Hamiltonian on various wave functions. To speed up the evaluation of first- and second-order derivatives, we use the FFTW (the Fastest Fourier Transform in the West) rather than the DVR-based matrix representation, thereby avoiding matrix operations altogether. This allows us to evaluate spatial derivatives with extremely high accuracy and with essentially the same speed as a multi-step finite difference formula. The time evolution of the TD-SLDA equations is performed using a multistep, fifth-order predictor-corrector-modifier Adams-Bashforth-Milne method:

$$p_{n+1} = \frac{y_n + y_{n-1}}{2} + \frac{h}{48}\left(119 y'_n - 99 y'_{n-1} + 69 y'_{n-2} - 17 y'_{n-3}\right),$$

$$m_{n+1} = p_{n+1} - \frac{161}{170}(p_n - c_n),$$

$$c_{n+1} = \frac{y_n + y_{n-1}}{2} + \frac{h}{48}\left(17 m'_n + 55 y'_n + 3 y'_{n-1} + y'_{n-2}\right),$$

$$y_{n+1} = c_{n+1} + \frac{9}{170}(p_{n+1} - c_{n+1}).$$

We have selected this method for its unique combination of high accuracy and numerical stability and economical function-evaluation footprint. It requires only two evaluations of the right hand side of the differential equations per time step, a number that cannot be reduced without going to a lower accuracy numerical method. In each time dependent simulation we store various time dependent densities on the lattice to be used for later studies. The time step is chosen so the relative truncation error in the Adams-Bashforth-Milne method is between $10^{-7}$ and $10^{-15}$. Simulation results for runs with up to $10^6$ time steps show that this method is very stable and numerically accurate.

The SLDA software is written primarily in C with support drivers in Fortran. We have tested and developed our software utilizing several compilers including those from GNU Compiler Class, Portland Group, Pathscale, Intel, AMD , Cray, and IBM. The software was prototyped in UNIX (Linux) operating environments. We have attempted to subscribe to the standards found in: ISO/IEC JTC1/SC22 (Programming languages and operating systems) C99 + TC1 + TC2, WG14 N1124, dated 2005-05-06 (http://www.open-std.org/JTC1/SC22/WG14/www/standards.html) ; F03 (08): P1(base), P2(variable length strings), P3(conditional compilation)

(http://www.nag.co.uk/sc22wg5 , http://www.co-array.org/caf intro.htm (co-arrays -remote memory ops)); POSIX Threads IEEE Std 1003.1, 2004 Edition;ISO/IEC standard 9945-1:1996 (http://www.unix.org/single unix specification). We use existing software libraries whenever possible. Here is a short list of library dependencies in no logical order: Message Passing Interface (MPI1,MPI2 specifications) (MPICH, Cray, Intel, OpenMPI), Basic Linear Algebra Subprograms (BLAS) (Automatically Tuned Linear Algebra Subroutines (ATLAS) , GOTO BLAS , Cray's Scientific Library (LIBSCI) , AMD's Core Math Library (ACML) , Intel's Math Kernel Library (MKL), IBM's Engineering and Scientific Subroutine Library (ESSL) ), POSIX Threads (PThreads), Open Multi-Processing (OpenMP), Linear Algebra PACKage (LAPACK) (ATLAS,LIBSCI, ACML, MKL, ESSL), Scalable LAPACK (ScaLAPACK) (Cray's LIB-SCI,Intel's CMKL, IBM's PESSL), Fastest Fourier Transform in the West (FFTW) , and LUSTRE (including liblut). For visualization we use VisIt and the Silo library. We have implemented segments of our time dependent codes with OpenMP compiler directives but have not been able to outperform the current MPI-only variant of our software. Also, we are actively prototyping a Pthread version of the unitary time-dependent superfluid gas code.

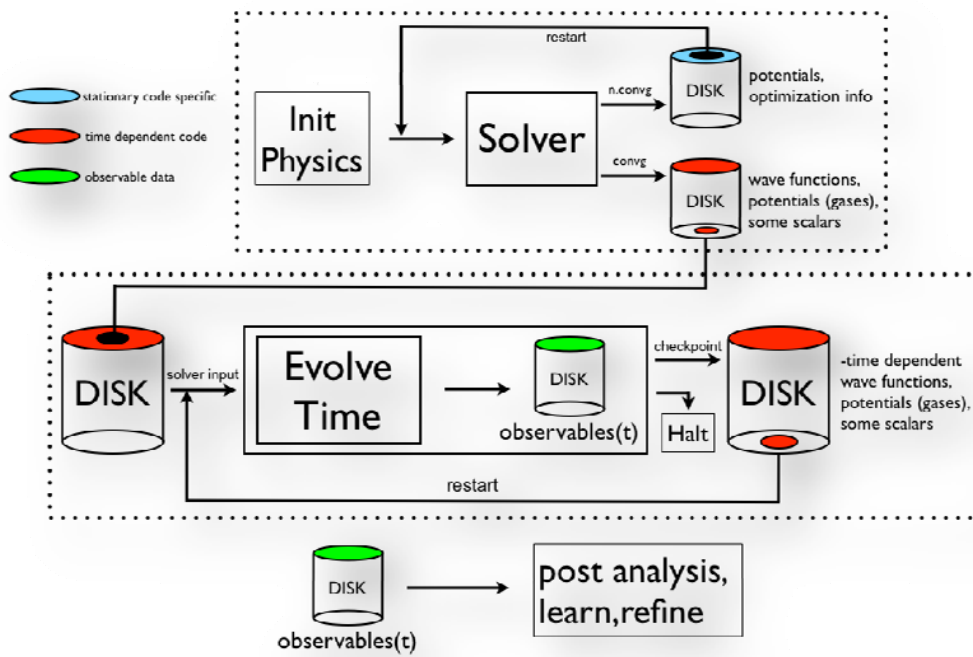

*Figure 8. Structure of the (TD)SLDA production software and workflow.*

The workflow of the SLDA software is depicted in Figure 8. These codes fall into the top dotted box in the flowchart. Related to the solvers are the respective time-dependent codes –shown in the other dotted box. When we make a numerical study we implement the Hamiltonian into the solver and self-consistently iterate until convergence. If, during a scheduled single execution event, the solver runs out of the requested time on the computer, the optimization details and potentials are restored from disk in a subsequent execution event and execution resumes where it was interrupted. This process continues until completion at which point the converged wavefunctions and potentials are stored in the Lustre file system. The time-evolution codes are initialized by reading the converged stationary solutions and potentials (and some relevant scalars). We typically adiabatically introduce the external probe that describes the experiment we are making in the time dependent code. The studies we have described require a very large number of time steps. We write time-dependent observable data at a given frequency of time steps. At the end of a single execution event, the time-dependent wave function data and a small number of potentials and scalars are stored during a checkpoint phase. In a subsequent run, a restart occurs from this data and continues in this manner until the simulation is deemed complete. The stored observable data is analyzed at a later time.

| 24408 | 48816 | 97632 | 195250 | PEs |
|---|---|---|---|---|
| 8 | 4 | 2 | 1 | WF/PE |
| 14.52 | 8.59 | 5.83 | 4.16 | T[s]/TS |
| 6.99E+14 | 8.72E+14 | 1.32E+15 | 2.02E+15 | INS/TS |
| 2.33E+14 | 2.53E+14 | 2.94E+14 | 3.74E+14 | FP_OP/TS |
|  | 84% | 62% | 44% |  |
| 0.76 | 0.80 | 0.89 | 0.96 | ins / cyc / pe |
| 0.33 | 0.29 | 0.22 | 0.19 | fp_op / ins |
| 3.00 | 3.44 | 4.50 | 5.40 | ()^-1 |

*Figure 8. $N_x=N_y=N_z=72$ TD-SLDA Strong Scaling Example. The floating point efficiency goes down as a function of increasing processes. However, a speed-up is measured in each case.*

By far the most expensive part of our studies are the time dependent codes. We are at the scale of the full machine today – the largest open-science supercomputer in the world at the times of these studies. For example, we have executed the time-dependent unitary gas code successfully on 217,752PEs ($62^3$) –over 97% of the full system JaguarPF and exhibiting nearly 64% ideal strong scaling over the same problem on half the PEs. The cost of executing a time step is the key performance metric. Below is a detailed example of a $72^3$ lattice strong scaling study. The percentages are fraction of ideal strong scaling achieved.

Here we present data from typical studies made with the SLDA software. The static codes exhibit the behavior that the solvers require a smaller resource, but the time evolution codes require a large resource and for a long time.

A small problem we have just completed as exemplary or our studies is a 5216 particle dilute gas in a trap on a 50x50x100 lattice with 1,039,17 2-component wavefunctions computed. Run 1 was the solver. It employed 51 parallel, parallel groups and 144 PEs / group and 129 iterations to converge. Run 2 is the first time dependent run. It reads the solutions from the solver, executes exactly one time step and writes one set of observables, and then checkpoints 8TB of data utilizing 24 Lustre io groups. (N.B. This run was made prior to improving our I/O capability over Lustre –see I/O for details) Run 3 restarts from disk, executes 2050 additional time steps, performs 25 additional observable I/O events, and exists cleanly. The cost of this example was around 635,000 CPU-Hours on the system. To complete the physical simulation would require another 100k time steps roughly leading to a cost of about 12M CPU-Hours total. The event is described in the following table:

| Machine Data | Run 1 | Run 2 | Run 3 | Total |
|---|---|---|---|---|
| Instructions | 3.335083409e17 | 3.538162667e18 | 4.415760819e18 | 8.287431827e18 |
| Floating Point Ops | 3.628450357e15 | 1.91882289e14 | 3.235240267e17 | 3.273443593e17 |
| Wall Time(s) | 11,085.29975 | 8,291.248311 | 12,913.89644 | 32,290.4445 |
| CPU $(hours) | 22,614.01149 | 239,333.7919 | 372,770.3823 | 634,718.1857 |
| PEs | 7,344 | 103,917 | 103,917 | |

We employ Lustre IO groups for the checkpoint restart phases of our codes. The scale of data moved to and from disk is 44 * NWF * $N_xN_yN_z$* sizeof( double complex ) BYTEs and recall that NWF is $O(N_xN_yN_z)$. For example, we achieve anywhere between 3 GBps and 20 GBps with our current implementation. On average, we achieve roughly 10 GBps for both reading and writing. The performance is a function of the state of the computer's shared communication and i/o network. We write about 10 densities at some frequency per simulation in a related simulation of a nuclear system and each of these has size $N_xN_yN_z$ * sizeof(double) BYTES –this is our data storage demand.